\documentclass{aa}  
\usepackage{graphicx}
\usepackage{txfonts}
\usepackage{textcomp}

\newcommand{\thco}{{\hbox {\ensuremath{\mathrm{^{13}CO}} }}}
\newcommand{\twco}{{\hbox {\ensuremath{\mathrm{^{12}CO}} }}}
\newcommand{\tastar}{\hbox {T$_{\rm A}^*$ }}
\newcommand{\kmps}{\ensuremath{\mathrm{km\,s^{-1}}}}
\newcommand{\twcop}{{\hbox {\ensuremath{\mathrm{^{12}CO}}}}}
\newcommand{\thcop}{{\hbox {\ensuremath{\mathrm{^{13}CO}}}}}

\begin{document}

\title{Radio observations of globulettes in the Carina nebula  \thanks{Based on observations collected with the Atacama Pathfinder Experiment (APEX), Llano Chajnantor, Chile (O-091.F-9316A and O-094.F-9312A)}}


  \author{L. K. Haikala\inst{1}
        \and G. F. Gahm\inst{2}
        \and T. Grenman\inst{3}
        \and M. M. M\"akel\"a\inst{4,5}
          \and C. M. Persson\inst{6}
                }


\institute{Universidad de Atacama, Copayapu 485, Copiapo, Chile \\
        email: \mbox{lauri.haikala@uda.cl}   
        \and Stockholm Observatory, AlbaNova University Centre, Stockholm University, SE-106 91 Stockholm, Sweden
        \and Applied Physics, Department of Engineering Sciences \& Mathematics, Lule{\aa} University of Technology, SE-97187 Lule\aa, Sweden
       \and Dr. Karl Remeis-Sternwarte, Astronomisches Institut der Universit\"at Erlangen-N\"urnberg, Sternwartstrasse 7, D-96049 Bamberg, Germany
              \and Department of Physics, Division of Geophysics and Astronomy, PO Box 64, FI-00014 University of Helsinki, Finland
         \and Chalmers University of Technology, Department of Earth and Space Sciences, Onsala Space Observatory, SE-43992 Onsala, Sweden }

   \date{}

 \abstract
   {The Carina nebula hosts a large number of globulettes. An optical study of these tiny molecular clouds shows that the majority are of planetary mass, but there are also those with masses of several tens up to a few hundred Jupiter masses.} 
   {We seek to search for, and hopefully detect, molecular line emission from some of the more massive objects; in case of successful detection we aim to map their motion in the Carina nebula complex and derive certain physical properties.}
   {We carried out radio observations of molecular line emission in $^{12}$CO and $^{13}$CO (2--1) and (3--2) of 12 globulettes in addition to positions in adjacent shell structures using APEX.}
   {All selected objects were detected with radial velocities shifted relative to the emission from related shell structures and background molecular clouds. Globulettes along the western part of an extended dust shell show a small spread in velocity with small velocity shifts relative to the shell. This system of globulettes and shell structures in the foreground of the bright nebulosity surrounding the cluster Trumpler 14 is expanding  with a few km\,s$^{-1}$ relative to the cluster. A couple of isolated globulettes in the area move at similar speed. Compared to similar studies of the molecular line emission from globulettes in the Rosette nebula, we find that the integrated line intensity ratios and line widths are very different. The results show that the Carina objects have a different density/temperature structure than those in the Rosette nebula. In comparison the apparent size of the Carina globulettes is smaller, owing to the larger distance, and the corresponding beam filling factors are small. For this reason we were unable to carry out a more detailed modelling of the structure of the Carina objects in the way as performed for the Rosette objects. }
  {The Carina globulettes observed are compact and denser than objects of similar mass in the Rosette nebula. The distribution and velocities of these globulettes suggest that they have originated from eroding shells and elephant trunks. Some globulettes in 
the Trumpler 14 region are quite isolated and located far from any shell structures. These objects move at a similar speed as the globulettes along the shell, suggesting that they once formed from cloud fragments related to the same foreground shell. }
  

 \maketitle
%

\section{Introduction}
\label{sec:intro}

The Carina nebula (NGC 3372), hereafter labelled CN, is a magnificent giant  \ion{H}{ii} region powered by radiation and winds from more than 70 O-type stars (see the review by Smith \& Brooks \cite{smith08}). Several young clusters (Trumpler 14, 15, and 16; Collinder 228 and 232; and Bochum 10 and 11) are located in this region. A large number of pre-main-sequence stars have been identified from optical, infrared, and X-ray surveys (e.g. Tapia et al. \cite{tapia03}; Ascenso et al. \cite{ascenso07}; Sanchawala et al. \cite{sanchawala07a}, \cite{sanchawala07b}; Povich et al. \cite{povich11}; Gaczkowski et al. \cite{gaczkowski13}; Preibisch et al. \cite{preibisch14}; Kumar et al. \cite{kumar14}; Beccari et al. \cite{beccari15}), and Zeidler et al. (\cite{zeidler16}) estimated the total number of pre-main-sequence objects in the region at 164~000. Hartigan et al. (\cite{hartigan15}) presented candidate jets emanating from young stars in an area covering more than a square degree of the nebula. 

Optical images show extended networks of dark, obscuring shells, and as pointed out in Smith et al. (\cite{smith00}, \cite{smith10a}, \cite{smith10b}), star formation proceeds in the remnant shells. The large scale distribution of dust in the CN complex can be overviewed in the maps based on surveys from the submm range (Preibisch et al. \cite{preibisch11}; Pekruhl et al. \cite{pekruhl13}) and the far-infrared (Preibisch et al. \cite{preibisch12}; Roccatagliata et al. \cite{roccatagliata13}). The complex is located close to the Galactic plane, but radio maps of molecular line emission from the area show that the dust shells associated with the nebula move at different speeds than the background emission from the Carina arm (Whiteoak et al. \cite{whiteoak84}; Grabelsky et al. \cite{grabelsky88}; Cox \& Bronfman \cite{cox95}; Brooks et al. \cite{brooks98}, \cite{brooks03}; Yonekura et al. \cite{yonekura05}).

\begin{table*}[!ht] 
\centering
\caption{Globulettes, fragments, and shell structures observed with APEX. }
\begin{tabular} {lccccc lccccc}
 \hline\hline
\noalign{\smallskip}
CN & R.A. & Dec. & $\alpha$ & $\beta$ & $\bar r $ & CN & R.A. & Dec. & $\alpha$ & $\beta$ & $\bar r $ \\
& (J2000.0) &(J2000.0) & (") &  (") &  (kAU) & & (J2000.0) &(J2000.0) & (") &  (") &  (kAU)   \\
\noalign{\smallskip}
\hline
 \\  
38  & 10:43:39.1 & -59:32:58 & 1.3 & 1.0        &   3.3 & 93-Frag  & 10:44:11.2 & -59:33:25  & 9.9   &  (*) &  (*)\\  
38-off1 &(-20,0) & & & &                                & 93-Frag-off1 & (-20,+10)    & & &  &                       \\
38-off2 &(+20,0) & & & &                                &  199 &  10:44:41.3 & -59:46:27     & 1.5 & 1.2 & 3.9        \\  
38-Shell1 & 10:43:41.0 & -59:33:42  & &        &     &   199-off1 & (+30,0)    & & &  &                 \\                             
38-Shell2 & 10:43:32.5 & -59:34:26  & &        &         & 199-off2 & (0,+30)    & & &  &                 \\
38-Shell3 & 10:43:28.1 & -59:34:12  & &        &        & 199-Shell1 & 10:44:42.4   & -59:48:30   &  &  &  \\
78  &  10:43:56.4 & -59:36:53 & 2.8 & 2.1   & 7.0 & 199-Shell2 &  10:45:01.3  & -59:47:03    &  &  &             \\
78-off1 & (0,-20)    & & &  &               &    $j$ & 10:45:01.3 & -60:02:11 & 3.9 & 3.2 &   10.3         \\          
78-Shell1  &  10:43:51.1 & -59:36:53 &  &   &           &  $j$-off1 & (0,+20)    & & &  &                  \\
79  & 10:43:56.4         & -59:39:03 & 1.2 & 1.2 &   3.5    & E  & 10:45:16.3 & -59:28:11 & 1.6 & 1.0 &   3.8 \\
79-off1 & (0,-30)    & & &  &               & E-off1 & (+20,-20)    & & &  &                  \\
80  & 10:43:56.6 & -59:36:33    & 2.9 & 2.3      & 7.6     &  F  & 10:45:18.6 & -59:28:01 & 3.0 & 2.6 &   8.1              \\
80-off1 & (+35,0)   & & &  &               & F-off1 & (0,-20)    & & &  &                    \\
93  &  10:43:59.6        & -59:32:38     & 1.7 & 1.3     &   4.4        & F-Shell1 &  10:45:21 & -59:27:36        &  &  &             \\
93-off1 & (-20,-20)    & & &  &             & HH 1006  & 10:46:33.0 & -60:03:54 & 4.5 & 2.0     &   9.4              \\
93-off2 & (0,-20)    & & &  &               & HH 1006-off1 & (0,+20)    & & &  &              \\
93-off3 & (+20,0)    & & &  &               & & & & &  &          \\
119 &  10:44:07.6 & -59:39:36   & 1.7 & 1.2 & 4.2  & & &  &      & &  \\
119-off1 & (+30,-20)    & & &  &          & & & &  & &     \\
119-off2 & (0,-30)    & & &  &  & & & &  &  &       \\
119-Shell1 &  10:44:04.1 & -59:40:01 &  &  & &  & & & & &  \\ 
\\    
\hline
(*) = irregular shape
\end{tabular}
\label{table:objects} 
\end{table*} 

There are also a large number of isolated dark clouds seen in silhouette against the bright nebulosity in CN. Some are large and have very irregular shapes, such as the ``Defiant Finger" described in Smith et al. \cite{smith04}, but there are also smaller fragments shaped like worms or long, narrow cylinders. Most of the dark clouds are of more regular shape. A number of small obscuring structures in CN were noted by Smith et al. (\cite{smith03}, \cite{smith04}) from images obtained with the Hubble Space Telescope (HST); these were regarded as possible proplyds. The study presented here is based on inventory by Grenman \& Gahm (\cite{grenman14}; hereafter called G14) of tiny, roundish, starless clouds, called {\it globulettes}, from HST H$\alpha$ images of CN. 

Small-sized dark clouds in \ion{H}{ii} regions were noted long ago by Bok \& Reilly (\cite{bok47}), Thackeray (\cite{thakeray50}), and Herbig (\cite{herbig74}). Later studies of such starless cloudlets by De Marco et al. (\cite{marco06}) and Gahm et al. (\cite{gahm07}) showed that most of the objects have radii $<$ 10~kAU with size distributions that peak at $\sim$~2.5~kAU. Gahm et al. (\cite{gahm07}) derived masses from extinction measures indicating that most objects have masses $<$~13~M$_{J}$ (Jupiter masses). Follow-up near-infrared (NIR) imaging and radio molecular line observations of the larger globulettes in the Rosette nebula, hereafter labelled RN, showed that these objects are dense and that the gas is molecular even close to the surface layers (Gahm et al. \cite{gahm13}; hereafter called G13; M\"akel\"a et al. \cite{makela14}). Mass estimates derived from modelling the molecular line emission (based on the gas content) were found to be similar, but systematically higher compared to those derived from extinction (based on the dust content). The system of globulettes in the northern part of RN moves outwards with velocities of about 22 km~s$^{-1}$ relative to the central cluster. The CN globulettes differ from those in RN and other regions in that they are on the whole smaller and denser as noted in G14. 

\begin{figure*}[t]
\centering
\includegraphics[angle=00, width=7cm]{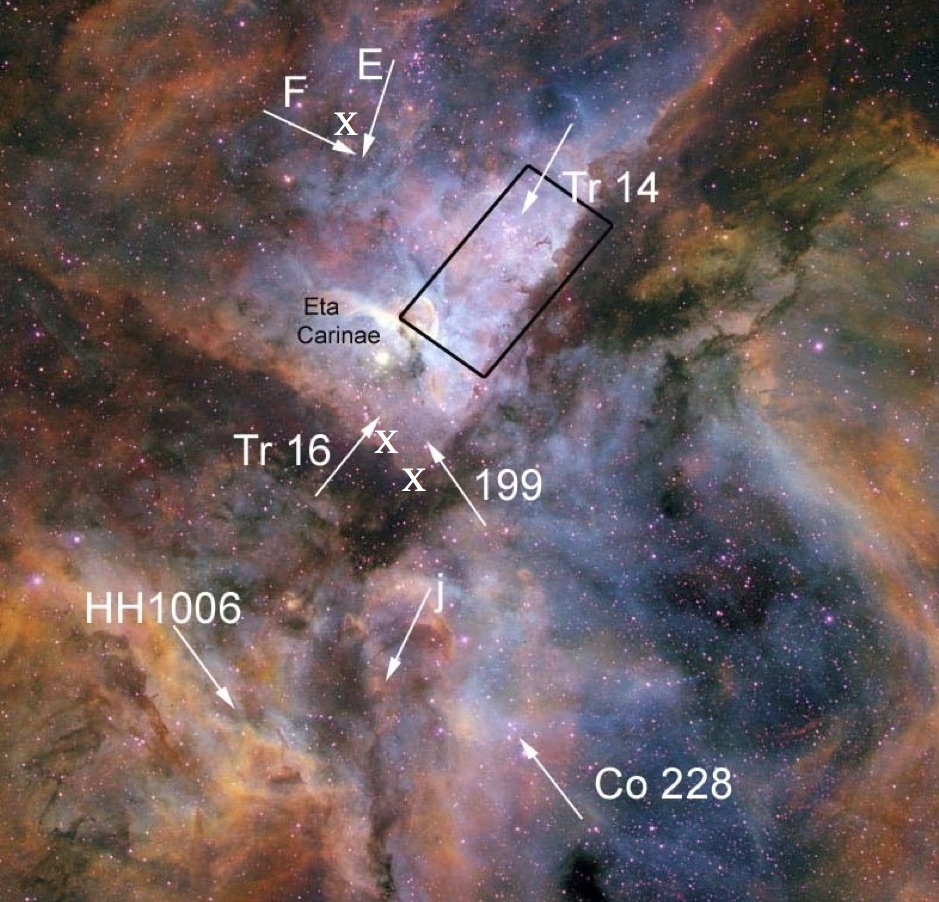}
\includegraphics[angle=00, width=10cm]{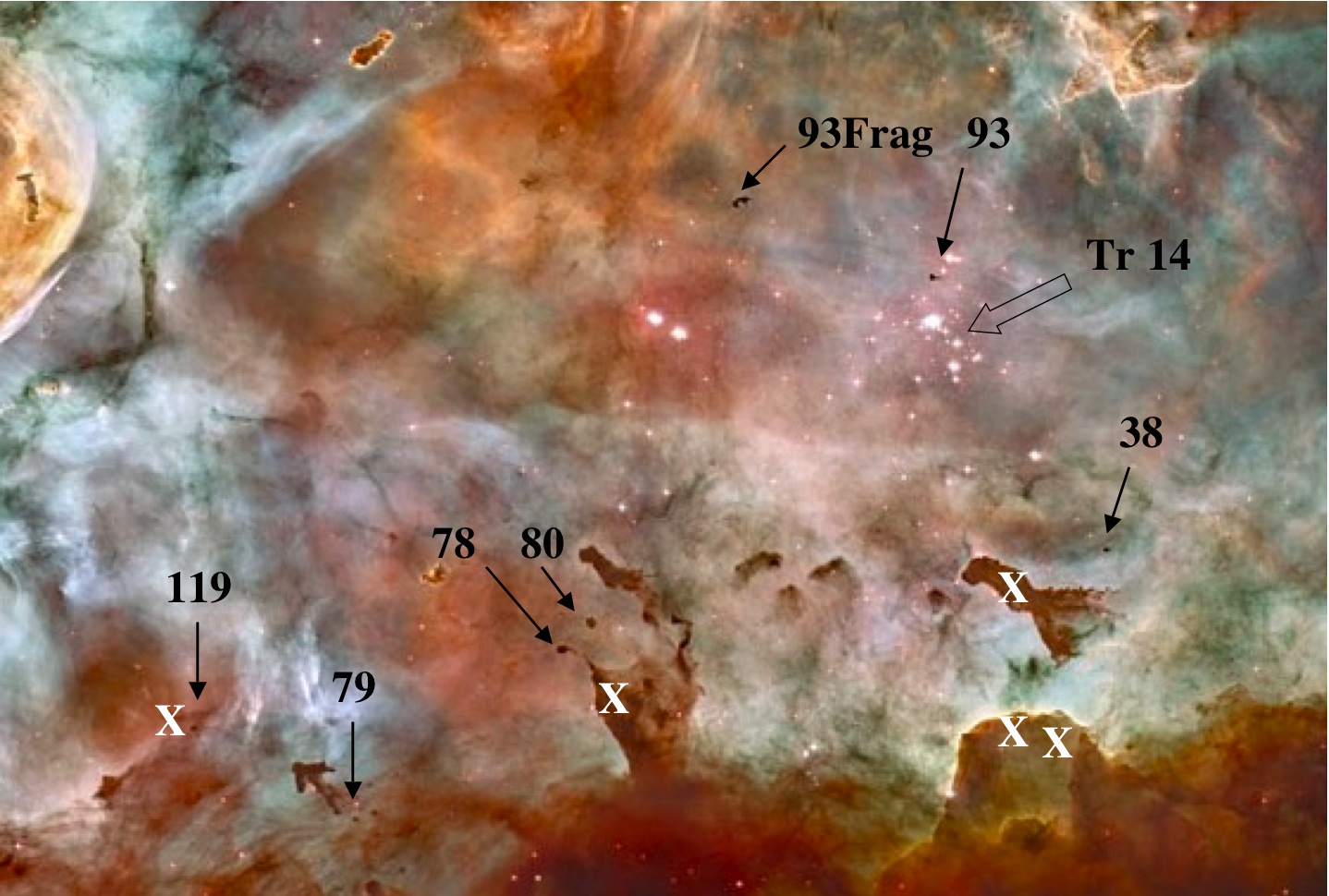}
\caption{Overview of observed positions. Left: the central region of the Carina nebula, where the locations of the star $\eta$~Carin{\ae} and three stellar clusters are indicated in addition to the four objects F, E, $j$, and HH 1006. The image spans about 1\degr $\times$ 1\degr. North is up and east is left (image: Nathan Smith, Univ. of Minnesota, NOAO, AURA, and NSF). The framed region is enlarged in the right panel, where observed objects are indicated with arrows. The image spans $\sim$~12\arcmin $\times$ 7\arcmin (image: Hubble Space Heritage Team). Positions in shell structures are denoted as white crosses in both images.} 
\label{fig:fields}
\end{figure*}

\begin{figure*}[t]
\centering
\includegraphics[angle=00, width=3.45cm]{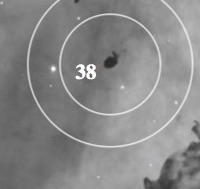}
\includegraphics[angle=00, width=3.43cm]{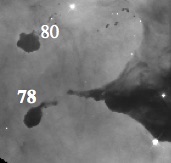}
\includegraphics[angle=00, width=3.51cm]{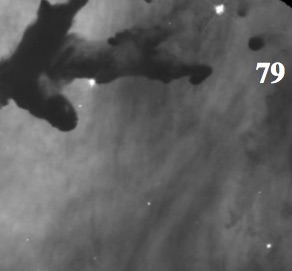}
\includegraphics[angle=00, width=3.64cm]{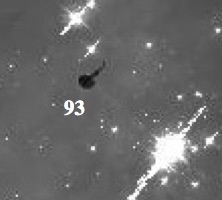}
\includegraphics[angle=00, width=3.49cm]{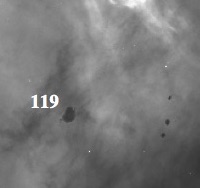}
\includegraphics[angle=00, width=3.44cm]{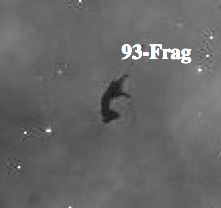}
\includegraphics[angle=00, width=3.49cm]{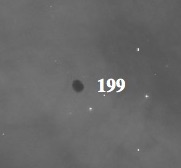}
\includegraphics[angle=00, width=3.46cm]{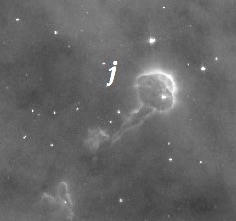}
\includegraphics[angle=00, width=3.65cm]{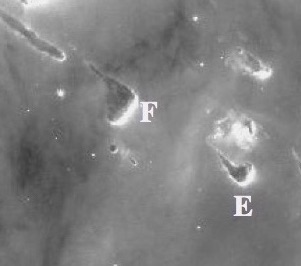}
\includegraphics[angle=00, width=3.48cm]{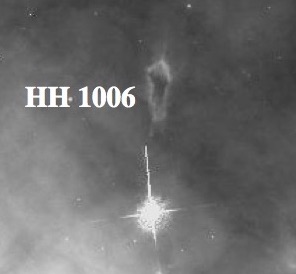}
\caption{HST H$\alpha$ images with observed objects marked. The circles, upper left, indicate the extensions of the beams (FWHM) in the 230 and 345 GHz bands. All images span about 40\arcsec $\times$ 40\arcsec. North is up and east is left.} 
\label{fig:images}
\end{figure*}

The question of whether or not globulettes may collapse to form free-floating low-mass objects, such as brown dwarfs or planetary mass objects, has been of special concern (Gahm et al. \cite{gahm07}, G13, and G14, and references therein). In the case of RN such objects shoot out in the interstellar environment at high speed. Numerical simulations in Kuutmann (\cite{kuu07}) predict lifetimes of $\sim$~$10^{4}$ years, increasing with mass, and it was found that larger globulettes may collapse before evaporation has proceeded very far. Haworth et al. (\cite{haworth15}) calculated the evolution of globulettes assuming pressure-confined isothermal Bonnor-Ebert spheres. Also in these model simulations the globulettes survive a long time, but do not contract into low-mass objects except in the case where they catch up with the expanding and retarding shell. However, from relative intensities of molecular lines in RN objects G13 found that the interior structure is very different from the Bonnor-Ebert case, and furthermore the NIR-imaging revealed very dense cores at least in some of the larger objects. 

In the present investigation we follow up the optical study in G14 of the CN globulettes with radio molecular line observations. In particular, we want to explore how the objects move relative to stellar clusters and shell structures in the area, derive certain physical properties of the objects, and to make comparisons with the RN globulettes investigated in G13. 

The paper is organized as follows. Observations and selected objects are presented in Section~\ref{sec:obs}. In Sect.~\ref{sec:results} we report on the results from our radio observations. The results are analysed and discussed further in Sect.~\ref{sec:disc}, and we conclude with a summary in Sect.~\ref{sec:conclusions}.

\section{Observations and objects}
\label{sec:obs}

\subsection{Radio observations}
\label{sec:radio}

Radio observations of the \mbox{$J$\,=\,3\,--\,2} and \mbox{$J$\,=\,2\,--\,1} transitions of \twco and \thco  were carried out in service mode in May 12-13, 2013, July 6-8, 2013, and August 15-20, 2014 with the 12~m APEX telescope at Llano Chajnantor, Chile: $^{12}$CO \mbox{$J$\,=\,3\,--\,2}  at 230.538~GHz, and \mbox{$J$\,=\,3\,--\,2} at 345.796~GHz, and $^{13}$CO \mbox{$J$\,=\,2\,--\,1} at 220.399~GHz, and \mbox{$J$\,=\,3\,--\,2} at 330.588~GHz.  We used two single sideband heterodyne SIS-receivers mounted on the Nasmyth-A focus: APEX-1 and APEX-2. All observations were performed in position-switching mode. The  telescope FWHM is 27$\arcsec$ at 230 GHz and 18$\arcsec$ at 345 GHz, and the corresponding main beam efficiencies are 0.75 and 0.73. The 32\,768 channel RPG eXtended bandwidth Fast Fourier Transform Spectrometer (XFFTS) with a bandwidth of 2.5 GHz was used. The channel spacing 76.3~kHz corresponds to velocity spacing $\Delta v$ of 0.1 and 0.07~km\,s$^{-1}$ at 230 and 345~GHz, respectively. The pointing was checked regularly and the error is estimated to be within 2$\arcsec$.

\subsection{HST images}
\label{sec:HST}

Optical images of the CN complex were downloaded from the HST archive, cycle 13 and 14 programs GO-10241 and 10475 (principal investigator N. Smith) based on observations with the ACS/WFI camera, which contains two CCDs of 2048~$\times$~4096 pixels glued together with a small gap in between. The pixel size corresponds to $\approx$~0.05$\arcsec$ pixel$^{-1}$ and the field of view is $202$\arcsec$ \times 202$$\arcsec$. All images selected were exposed for 1000~s through the narrowband filter F658N, covering the nebular emission lines of H$\alpha$ and [\ion{N}{ii}].

\subsection{Selected objects}
\label{sec:objects}

The majority of the globulettes listed in the survey in G14 are very tiny, and we selected a sample of the most massive objects that could hopefully be detected with APEX.  In addition, we observed several positions in shell structures or trunks in the vicinity of the selected globulettes. Finally, we selected some objects that were not regarded as typical globulettes in G14. One is 93-Frag, a very irregular fragment, which, similar to CN~93, is located far from surrounding shell structures. Objects $j$, F, and E were noted in G14 but not included as regular globulettes. We also observed HH 1006, an elongated fragment hosting a collimated bipolar jet (Smith \cite{smith10a}). This object was observed before in different molecular transitions by Sahai et al. (\cite{sahai12}), who recognized an IR star within the globule as the driving source of the outflow. The embedded star is of intermediate mass according to Mesa-Delgado et al. (\cite{mes16}) and Reiter et al. (\cite{reiter16}). In the following we refer to all objects as globulettes.

The observed targets with designations from G14 are listed in Table~\ref{table:objects} with positions in Columns 2 and 3 with associated offset positions in R.A. and Dec. in parenthesis in Column 2. In Columns 4 and 5 we list the semi-major and semi-minor axes in arcseconds as given in G14 and as measured for the added objects. The average radii, expressed in kAU, are in Column 6. The dimensions are based on a distance to the CN complex of 2.9 kpc as derived in Hur et al. (\cite{hur12}), but we note that a distance of 2.3 kpc has often been assumed (Smith \cite{smith08}).    

The positions of observed objects can be overviewed in Fig.~\ref{fig:fields}. The left panel covers a large area of the brightest part of the nebula with the star $\eta$ Carinae in the middle. The positions of the young stellar clusters Tr 14, 16, and Co 228, and the four objects F, E, $j,$ and HH 1006 are indicated together with some shell positions (crosses). Most of our targets fall along the western part of the characteristic V-shaped dust shell within the framed area, which is enlarged in the right panel. The positions of observed globulettes are denoted with arrows and designations from Table~\ref{table:objects}, and white crosses indicate observed shell positions. Close-ups of the regions containing selected objects are shown in the HST H$\alpha$ images in Fig.~\ref{fig:images}.

\section{Results}
\label{sec:results}

\begin{figure}[t]
\centering
\includegraphics[angle=00, width=8.5cm]{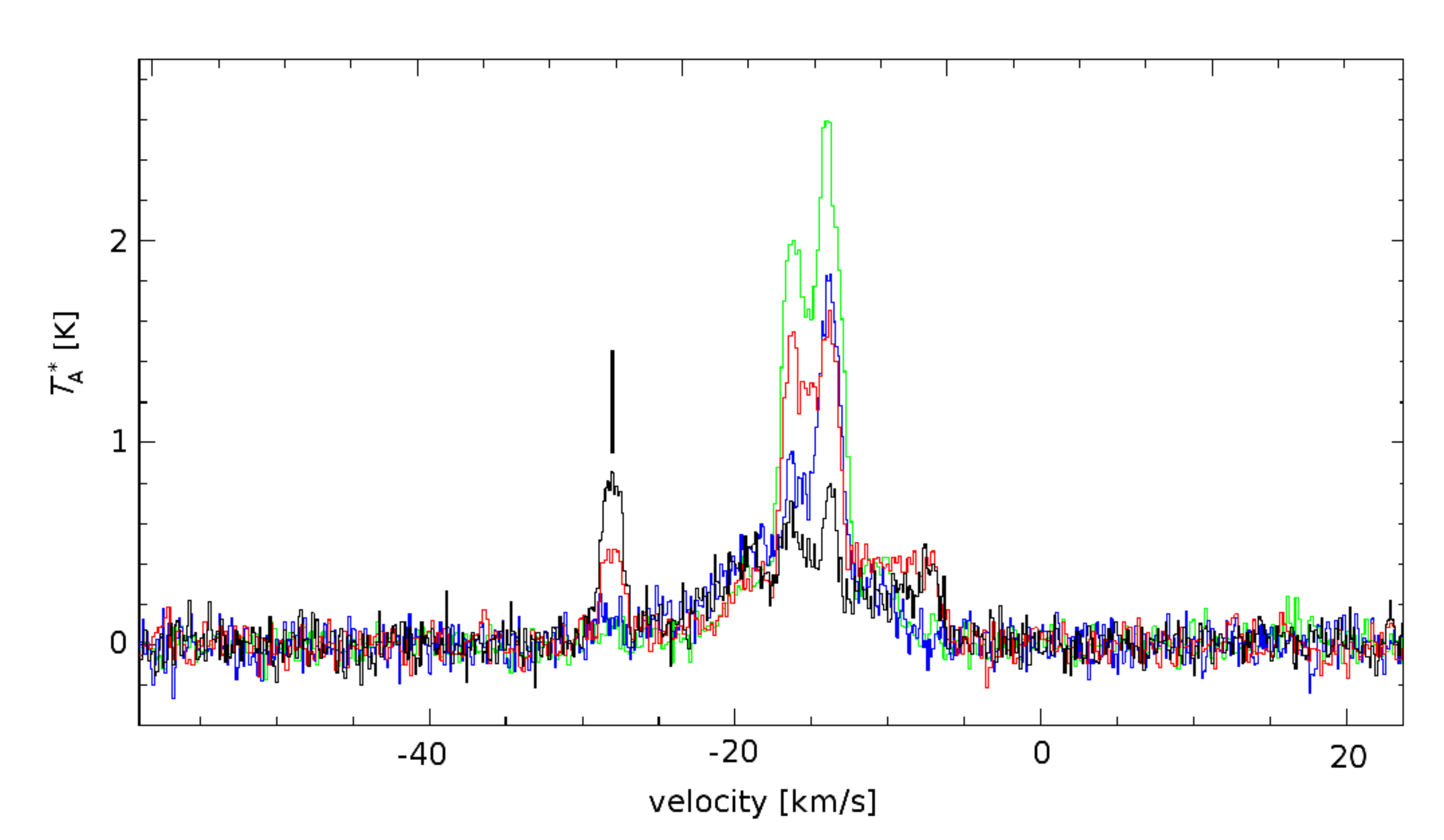}
\caption{ $^{12}$CO(3--2) and (2--1) Hanning-smoothed spectra of CN 93 are plotted in black and red, respectively. The  (3--2) and (2--1) spectra of the corresponding off-position are plotted in blue and green. Strong background components are present at velocities from -5 to -20 km s$^{-1}$. The signal at -28 km~s$^{-1}$ is related to the globulette (marked). Velocities are relative local standard of rest (LSR).} 
\label{fig:cn93}
\end{figure}

\begin{figure}[t]
\centering
\includegraphics[angle=00, width=3.5cm]{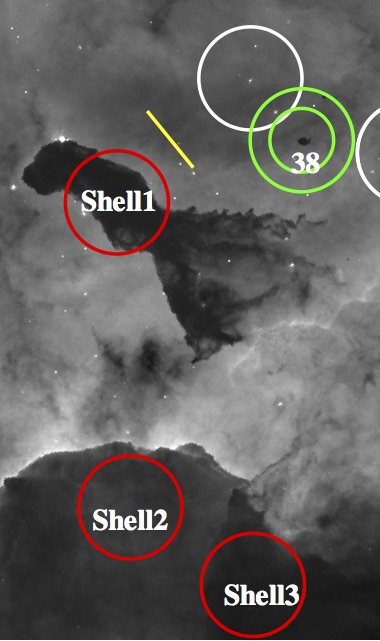}
\includegraphics[angle=00, width=5.3cm]{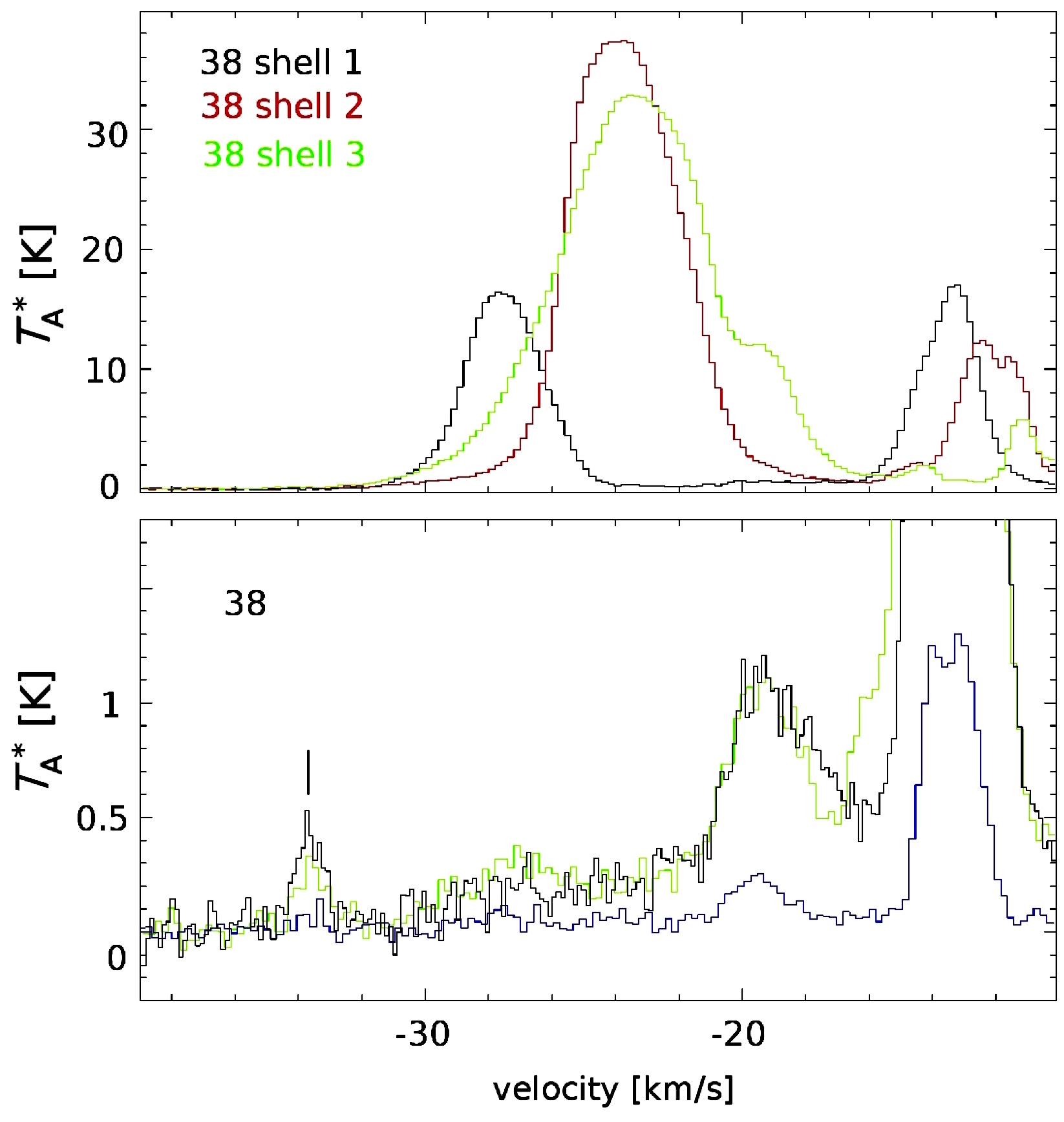}
\caption{Left: Positions observed in the area surrounding CN~38. Green circles at the on-position indicate the beam sizes at 230 GHz and 345 GHz. Shell positions are indicated in red and off-positions are shown in white. Image size is $\sim$45\arcsec$\times$120\arcsec. The yellow bar shows the orientation of the east-west direction (east left). Right panels: The corresponding Hanning-smoothed shell spectra (top) and the on-position spectra of CN~38 (below); green: $^{12}$CO(2--1), black: $^{12}$CO(3--2), and blue: $^{13}$CO(2--1).} 
\label{fig:38region}
\end{figure}

\begin{figure}[t]
\centering
\includegraphics[angle=00, width=3.75cm]{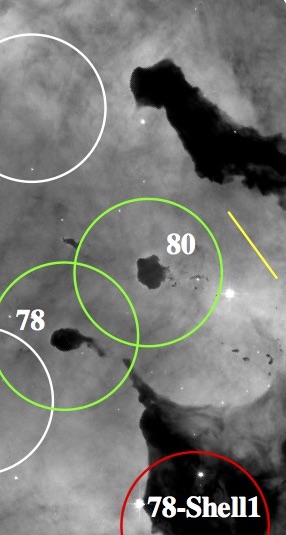}
\includegraphics[angle=00, width=4.65cm]{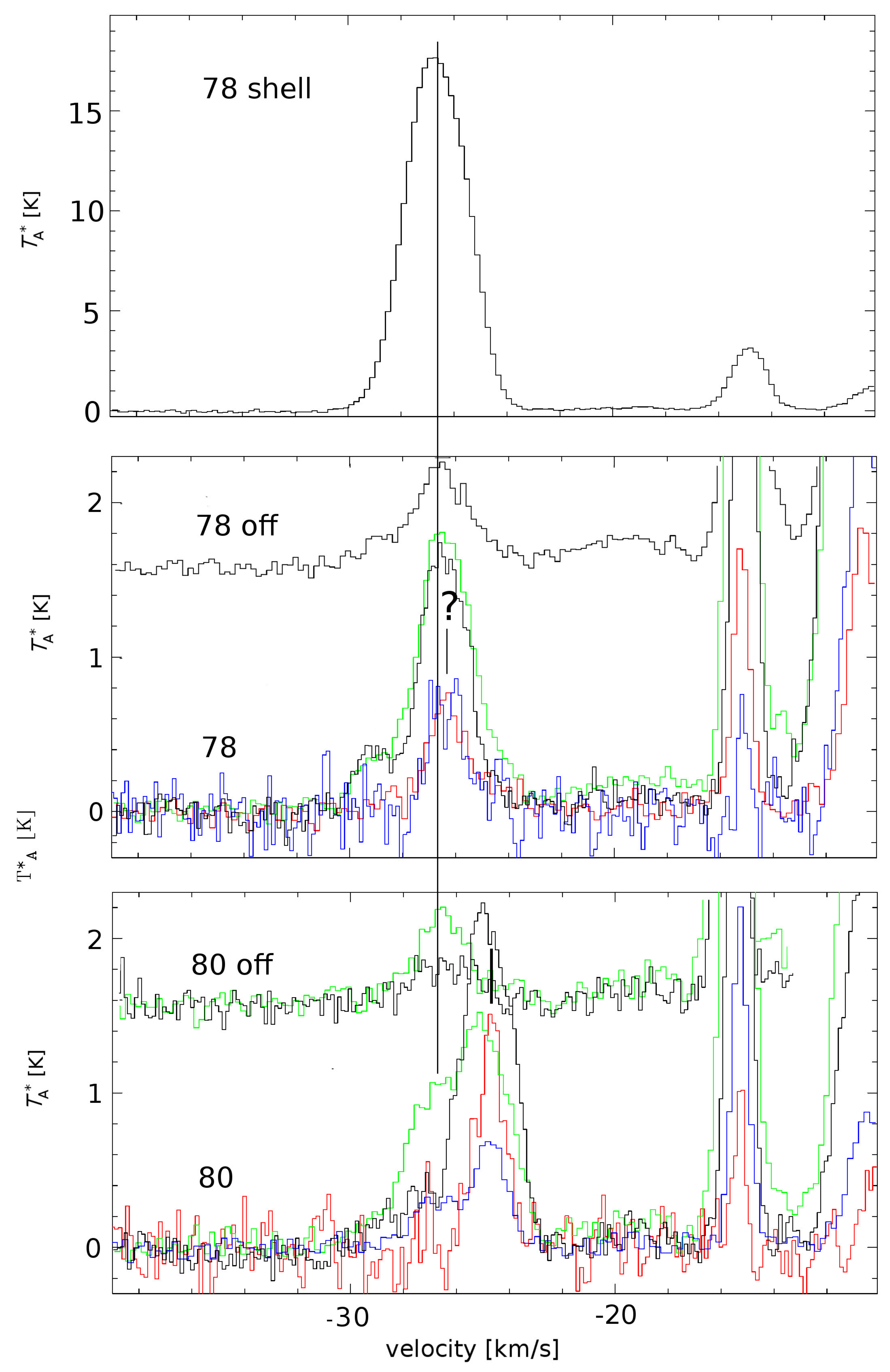}
\caption{Left: Observed positions in the region containing CN\,78 and 80 (circles correspond to the 230 GHz beam and colours as in Fig.~\ref{fig:38region}). The image size is $\sim$50\arcsec$\times$ 100\arcsec\  and the yellow bar shows the east-west direction (east left). Right: At the top is the Hanning-smoothed $^{12}$CO(2--1) spectrum of the shell (position 78-Shell1) and below the on- and off-position Hanning-smoothed spectra of CN\,78 and CN\,80 with an off-set of 1.6\,K. The signal  from CN~78 (at the question mark) is blended with shell emission, while CN~80 (marked) is red-shifted relative to the shell component (long vertical line). Colours as in Fig.~\ref{fig:38region} and with $^{13}$CO(3--2) in red.}
\label{fig:78-80region}
\end{figure}

\begin{figure*}[t]
\centering
\includegraphics[angle=00, width=18.0cm]{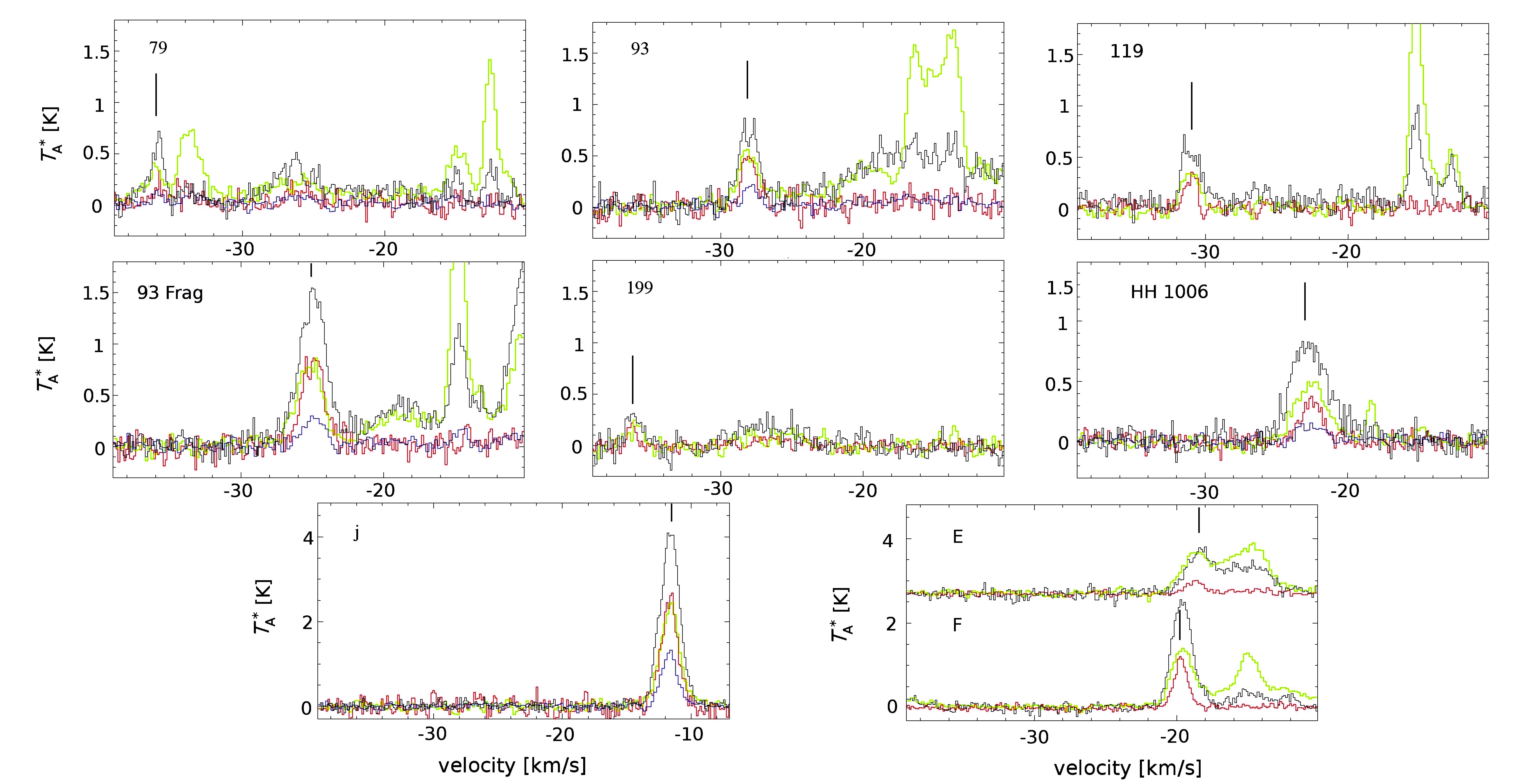}
\caption{Hanning-smoothed spectra from positions centred at selected globulettes;  green:$^{12}$CO(2--1); black: $^{12}$CO(3--2); blue: $^{13}$CO(2--1); red: $^{13}$CO(3--2). The signals related to the globulettes are marked. Corresponding spectra of CN~38, 78 and 80 are shown in Figs.~\ref{fig:38region} and \ref{fig:78-80region}. Strong emission from shells and/or background components are present in the panels of CN~79, 93, 119, 93-Frag, and objects E and F.} 
\label{fig:spectra}
\end{figure*}

Molecular line emission was detected from the positions of all objects. Very strong and broad lines are present in the velocity range -20 to +10 km s$^{-1}$ at most positions.{We identify these components} with shells on the remote side of the CN complex or more distant clouds in the Carina arm. The signals related to globulettes are at more negative velocities and with line widths of around 1.5 km\,s$^{-1}$ in the $^{12}$CO (3--2) transition. Fig.~\ref{fig:cn93} shows the on- and off-spectra of $^{12}$CO(2--1) and (3--2) for the isolated object CN~93, where the signal at -28 km s$^{-1}$ comes from the globulette. The intensity scale is expressed in terms of the antenna temperature reduced to outside the atmosphere, $T_\mathrm{A}^*$, and velocities refer to the local standard of rest (LSR) in all spectral diagrams.

Our spectra from positions in the foreground molecular shells show that these structures move at significantly more negative velocities than the background molecular gas. This is illustrated in Fig.~\ref{fig:38region}, which  shows spectra obtained at different locations in the area containing CN~38. Positions 38-Shell2 and 38-Shell3 (see Table~\ref{table:objects}) are located in the western part of the V-shaped shell and move at $\sim$ -24 km\,s$^{-1}$. Shell1 is located in what looks like a detached elephant trunk or with its root covered by foreground emission, moving at a more negative speed, -28 km\,s$^{-1}$. The globulette CN~38 is found at an even more negative velocity of -33.6 km\,s$^{-1}$ (lower panel).
 
Identification of the globulette signal is not straight forward in three cases: CN\,78, CN\,80, and object $E$.  The signals from CN~78 and 80 are mixed with those of the foreground shell structures as shown in Fig.~\ref{fig:78-80region}.  Comparing on- and off-spectra we identify the component at -24.8\,km\,s$^{-1}$ with  CN\,80. The component at  -26 km\,s$^{-1}$  is related to the base of the trunk-like feature with finger-like extensions shown in Fig.~\ref{fig:78-80region}. We suspect the signal from CN\,78 blends in velocity with the -26 km\,s$^{-1}$ component coming from the shell. We exclude the possibility that the CN\,78 line is the weak component at -29\,km\,s$^{-1}$ because the CO(3-2) and (2-1) lines are of the same intensity; the latter should be weaker because of beam dilution. Also owing to the size and mass of the globulettes, their signal is expected to be stronger than the -29 km\,s$^{-1}$ component. The identification of CN\,80 is based on the velocity shift compared to the signal from the shell, i.e. the strong  \twco and \thco (3--2) signals. Also the observed  $^{12}$CO(3--2)/$^{12}$CO(2--1) and $^{13}$CO(3--2)/$^{13}$CO(2--1) line intensity ratios point to a small size source. Emission at this velocity is seen neither in the off-position nor in the spectrum of the nearby CN\,78.  Similar deductions can be used to identify the signal from object E. Even though the signal from the globulette can be identified, its exact velocity, the  line half width, and $\tastar$ are uncertain because the line is blended with the signals from object F and the shell.

Spectra obtained from on-positions of all objects, excluding those shown in Figs.~\ref{fig:38region} and \ref{fig:78-80region}, are shown in Fig.~\ref{fig:spectra}. We regard objects CN~38, 80, 93, 119, 93-Frag, 199, HH 1006, and object $j$ as clearly detected. At the position of CN~79 several weaker signals are present, and we expect that emission from a large arrow-shaped structure to the east of CN~79 to enter the beam. 

In Table~\ref{table:spectra} we present line parameters derived for all objects and transitions. The optically derived mass as derived in G14 is listed in Column~2 expressed in Jupiter masses. For the objects not included in that survey, i.e. E, F, 93-Frag, and $j$, we derived masses with the same method as in G14. The globulettes are as a rule very dark and well confined, but H$\alpha$ emission from bright rims extends over objects $j$, F, and E, for which the optically determined masses should be regarded as lower limits (marked with an asterisk in the table). These objects show relatively strong lines indicating that they are more massive than estimated from optical extinction measures (see Sect.\ref{sec:lines}).The mass for HH~1006 was taken from Sahai et al. (\cite{sahai12}). Columns~3 to 7 give integrated line intensities and widths for different transitions in $^{12}$CO and  $^{13}$CO (left blank if not observed). Emission from CN~78 is present in $^{12}$CO(2--1) and (3-2), but because it is blended with emission from the shell at the same velocity we cannot give precise measures on intensities. Objects that were not clearly detected at specific transitions are indicated with a line. Column~8 gives the velocity (LSR) as measured from the $^{12}$CO(2--1) line. Very strong $^{12}$CO(2--1) emission was observed from shell-positions. Column~9 contains remarks, where objects with bright rims and pronounced tails are marked BR and T, respectively.

\begin{table*}
\centering
\caption{Data and line parameters for observed globulettes and shells. }
\begin{tabular} {lll ll llll  }
\hline\hline 
&&&\multicolumn{2}{c} {$^{12}$CO}&  \multicolumn{2}{c} {$^{13}$CO}    \\    
CN &  Mass &Transition& $\int{T^{\,*}_\mathrm{A}  \mathrm{d} v}$\tablefootmark{a} & $\Delta v$\tablefootmark{b}&   $\int{T^{\,*}_\mathrm{A}  \mathrm{d} v}$\tablefootmark{a} & $\Delta v$\tablefootmark{b} & $v_\mathrm{LSR}$\tablefootmark{c}& Remarks\\

&     ($M_{J}$)     &       & (K km s$^{-1}$)  & (km s$^{-1}$) &   (K km s$^{-1}$)  & (km s$^{-1}$)  & (km s$^{-1}$)  &\\
     \noalign{\smallskip}
     \hline
  &         &   &       &        &      &             &  &  \\
38       &  34  & 2--1 &   0.3   &   1.0  & --   & --   & -33.6    &  \\  
         &      & 3--2 &    0.5  &   1.1  &      &      &          &  \\
38-Shell1&      & 2--1 &   50.5  &   2.9  &      &      & -27.5    & in trunk \\  
38-Shell2&      & 2--1 &  164.9  &   4.0  &      &      &  -23.7   &  \\  
38-Shell3&      & 2--1 &  198.1  &   5.7  &      &      &  -23.4   &  \\
         &      & 2--1 &    8.0  &   1.4  &      &      &  -18.9   &  \\
78       & 128  & 2--1 &    blend  &   blend  & --   &  --  &  -26  & T \\  
         &      & 3--2 &    blend &   blend  & --   & --        &          &  \\78-Shell1&      & 2--1 &   52.2  &   2.7  &      &      &  -26.7   &  in trunk \\
79       &  29  & 2--1 &    0.4  &   1.5  &  0.1 &  0.8 &  -35.9   & \\ 
         &      & 3--2 &    0.8  &   1.2  &  0.3 &  1.8 &          &  \\
80       & 132  & 2--1 &    3.7  &   2.4  &  1.0 &  1.4 &  -24.6   &  \\  
         &      & 3--2 &    4.5  &   2.1  &  1.9 &  1.3 &          &  \\
93       &  63  & 2--1 &    1.1  &   2.0  &  0.3 &  1.1 &  -28.0   & T, projected against Tr 14 \\  
         &      & 3--2 &    1.3  &   1.5  &  0.7 &  1.2 &          &  \\
119      & 33  & 2--1 &    0.5  &   1.4  &      &      &  -31.0   & \\            &      & 3--2 &    1.0  &   1.7  &  0.3 &  0.9 &          &  \\
119-Shell1&     & 2--1 &    0.5  &   0.9  &      &      &  -31.9   &  \\  
93-Frag  & 208  & 2--1 &    1.9  &   2.3  &  0.5 &  1.7 &  -25.2   & irregular shape \\  
         &      & 3--2 &    3.6  &   2.4  &  1.7 &  1.9 &          &  \\
199      &  33  & 2--1 &    0.3  &   1.3  &      &      &  -36.0   &  \\  
         &      & 3--2 &    0.4  &   1.3  &  0.1 & 0.6  &          &  \\
199-Shell1&     & 2--1 &    9.9  &   1.5  &      &      &  -27.7   &  \\  
199-Shell2&     & 2--1 &   66.3  &   3.2  &      &      &  -28.0   &  \\
j        & 75*  & 2--1 &   4.4   &   1.8  &  2.0 &  1.5 &  -11.6   & T, BR  \\ 
&               & 3--2 &   7.8   &   1.8  &  4.1 &  1.5 &          &  \\  
E        & 20*  & 2--1 &   2.4   &   2.5  &  0.4 &  1.4 &  -18.7   & T, BR, identification unclear \\  
         &      & 3--2 &   2.1   &   2.1  &  0.9 &  1.4 &          &  \\  
F        & 65*  & 2--1 &   2.2   &   1.7  &      &      &  -19.6   & T,  BR, identification unclear \\  
         &      & 3--2 &   4.5   &   1.7  &  1.4 &  1.1 &          &  \\
F-Shell1 &      & 2--1 &  16.0   &   2.2  &      &      &  -15.3   &  \\ HH 1006  & 350** & 2--1 &   1.4   &   2.7  &  0.3 &  2.2 &  -22.5   & BR, jet/outflow  \\   
         &      & 3--2 &   2.1   &   3.1  &  0.6 &  1.8 &          &  \\ 
\hline
\label{Table: transitions}
\end{tabular}
\tablefoot{
\tablefoottext{a}{Integrated intensity.}
\tablefoottext{b}{Line width from a Gaussian fit.}
\tablefoottext{c}{LSR velocity of the globulette.}
\tablefoottext{-}{Not detected.}
\tablefoottext{*}{Lower limit.}
\tablefoottext{**}{Mass from Sahai et al. (\cite{sahai12})}
}
\label{table:spectra} 
\end{table*} 

As pointed out in G14 the CN globulettes are on the whole much smaller and denser than those found in other \ion{H}{ii} regions. Compared to our previous survey of globulettes in RN, at a distance of 1.5 kpc, the CN objects are also less extended for a given mass than in the RN survey, and therefore the beam filling is smaller. Nevertheless, we can claim detections of objects with masses as low as 30 $M_{J}$.

\section{Discussion}
\label{sec:disc}
 
\subsection{Velocity pattern}
\label{sec:velocity}

The globulettes shown in the right panel of Fig.~\ref{fig:fields} all move with about the same radial velocities with a small spread around a mean of -28 km\,s$^{-1}$. This is similar to the velocities of $\sim$~-27.5 km\,s$^{-1}$ obtained for the positions in the westward extension of the V-shaped dust shell. The close association between the globulettes and the shell, both in terms of velocity and position relative to trunk-like structures in the shell, indicates that they have formed from clumps in the shell that became detached at an earlier phase. Similar systems of detached globulettes were noted also in RN (G13). 

The two objects CN~93 and 93-Frag are quite isolated at projected distances of about 3 pc from the western shell. These objects share the velocity of the system of globulettes along the shell and could be leftovers from larger blocks in the same shell that underwent heavy erosion. Examples of eroding irregular blocks were given in G14; for example  one such block is called Fragment~4, which is also nearby in the sky to Tr 14. The globulettes CN~38, 199, j, E, and F move at slightly higher negative velocities than the adjacent shell structure, which was also found to be the case for several globulettes in RN (G 13). Either these objects have been accelerated through the interaction with stellar light since detachment or the shell has had time to slow down. 

Previous mappings of molecular line emission from the areas studied here have been rather coarse, but we note that Brooks et al. (\cite{brooks98}, \cite{brooks03}) recognized a molecular concentration inside the western dust lane with one velocity component at $\sim$ -25 km\,s$^{-1}$. This dust lane and associated globulettes are silhouetted against the nebulosity surrounding Tr 14, which is located about 3$\arcmin$ from the rim of the shell. It has been natural to assume that the excavation of the western dust lane and associated PDR regions along its edge are dominated by the interaction posed by Tr 14 (Brooks et al. \cite{brooks98}; Smith \& Brooks \cite{smith07}).    

Estimates of the mean heliocentric radial velocity of Tr 14 are based on spectroscopy of just a few stars and differ a great deal, ranging from +2.8 km\,s$^{-1}$ (Penny et al. \cite{penny93}) to -29 km\,s$^{-1}$ (Levato et al. \cite{levato91}). In more recent compilations Tr 14 is listed with -6.7 km\,s$^{-1}$ (Dias et al. \cite{dias02}) and  -15.0 km\,s$^{-1}$ (Kharchenko et al. \cite{kharchenko05}) based on three stars and one star, respectively. A straight mean of these results in a mean cluster velocity of -22.4 km\,s$^{-1}$. Although uncertain, it appears as the system of shells and associated globulettes along the western dust lane are blue-shifted by a few km\,s$^{-1}$ relative to Tr 14.  Expansion velocities of this order are expected from numerical simulations of expanding \ion{H}{ii} regions (e.g. Hosokawa \& Inutsuka \cite{hosokawa06}; Arthur et al. {\cite{arthur11}). We note that the northern system of shells and globulettes in RN is expanding at a much higher speed (G13). 

However, the western clouds are also exposed to UV radiation from the stars in the central region where $\eta$ Carinae and the rich cluster Tr 16 are located. As demonstrated in G14, elongated globulettes above and along the western shell have position angles pointing, not at Tr 14, but at the central region. A striking example of this is CN~93 (Fig.~\ref{fig:images}) with its bright-rimmed head oriented in the opposite direction of Tr 14. The elephant trunks along the western shell also bend in the direction of Tr 16. The two clusters are at the same distance according to Massey \& Johnson (\cite{massey93}) and Hur et al.(\cite{hur12}). It appears that the western system of globulettes and trunks have been sculpted not only from interactions caused by stars in Tr 14 but also from stars located further away in the direction of Tr 16. 

For the objects marked in the left panel of Fig.~\ref{fig:fields} we found a larger spread in radial velocity. CN~199, just above the edge of the V-shaped dust lane, moves at -36 km\,s$^{-1}$, which is close to the mean velocity of about -35 km\,s$^{-1}$ (LSR) measured for the members in the cluster Tr 16 in the same area (Dias et al. \cite{dias02}; Kharchenko et al. \cite{kharchenko05}). Object $j$, with a radial velocity of -12 km\,s$^{-1}$, is closer in the sky to Collinder 228 with a mean velocity of about -25 km\,s$^{-1}$ according to the same sources. Hence, this globulette appears to move in the opposite direction as the cluster. As noted above, HH 1006 hosts an embedded source driving a jet. The object moves at -23 km\,s$^{-1}$ and has line widths of $\sim$~3.2 km\,s$^{-1}$ (FWHM) in both $^{12}$CO lines. The lines are broader than for the other globulettes, which could be related to outflowing gas but with relatively small relative velocities in the line of sight (see also Sahai et al. \cite{sahai12}). A near-IR image of the [Fe II] emission from the jet extending north obtained by Reiter et al. (\cite{reiter16}) suggests that the northern outflow should be blue-shifted.

\subsection{Analysis of the molecular line emission}
\label{sec:lines}

The small size of the CN globulettes makes the interpretation of the molecular line data difficult. The mean diameter of the observed Carina globulettes is 4\farcs1 $\pm$ 1\farcs8 (median 3\farcs0) or 11.9 kAU $\pm$ 5.2 kAU with a median at 8.8 kAU (with a distance of 2.9 kpc). Molecular line observations using APEX of globulettes in the Rosette nebula were presented in G13 and these observations have an average diameter of 14\farcs7 $\pm$ 7\farcs7 or 20.0 kAU $\pm$ 11.8 kAU. Hence, the RN globulettes are not only apparently but also physically more extended than those in Carina. The beam filling factor corresponding to the average size of the RN objects is 0.34 in the APEX 346\,GHz 18\arcsec\ beam, but much smaller, only 0.04, for the CN objects. The ratio of the filling factors at the two frequencies remains the same, however. Because of the small filling factor we cannot make a detailed modelling of the physical structure of the CN globulettes as performed for the RN globulettes.

One unexpected finding in G13 concerning the globulettes in RN was that
the $^{12}$CO(3--2)/$^{12}$CO(2--1) and $^{13}$CO(3--2)/$^{13}$CO(2--1) line intensity ratios
do not depend on the source size but are close to unity. The $^{12}$CO(3--2)/$^{13}$CO(3--2) and $^{12}$CO(2--1))/$^{13}$CO(2--1) ratios are approximately 1/3 irrespective of the source size. As the APEX beam is 27\arcsec at 230 GHz and 18\arcsec\ at 345 GHz the beam filling factor at the higher frequency is much higher if the emission traces the same area. The beam filling factor depends strongly on the object size, and therefore the CO(3--2)/CO(2--1) ratios should increase as the source gets smaller. This indicates that for the RN globulettes the CO(3--2) and CO(2--1) emission does not trace the same region and that the CO (2--1) emission is more extended than that from the (3--2) transition. For the CN globulettes the $^{12}$CO(3--2)/$^{13}$CO(3--2) and $^{12}$CO(2--1))/$^{13}$CO(2--1) ratios are 1.6 $\pm$ 0.3 and 2.1 $\pm$ 0.5, respectively, and the integrated intensity of the (3--2) lines dominate over the (2--1) lines unlike for the RN objects, where they are similar.

\begin{figure}[h]
\centering
\includegraphics[angle=00, width=8.8cm]{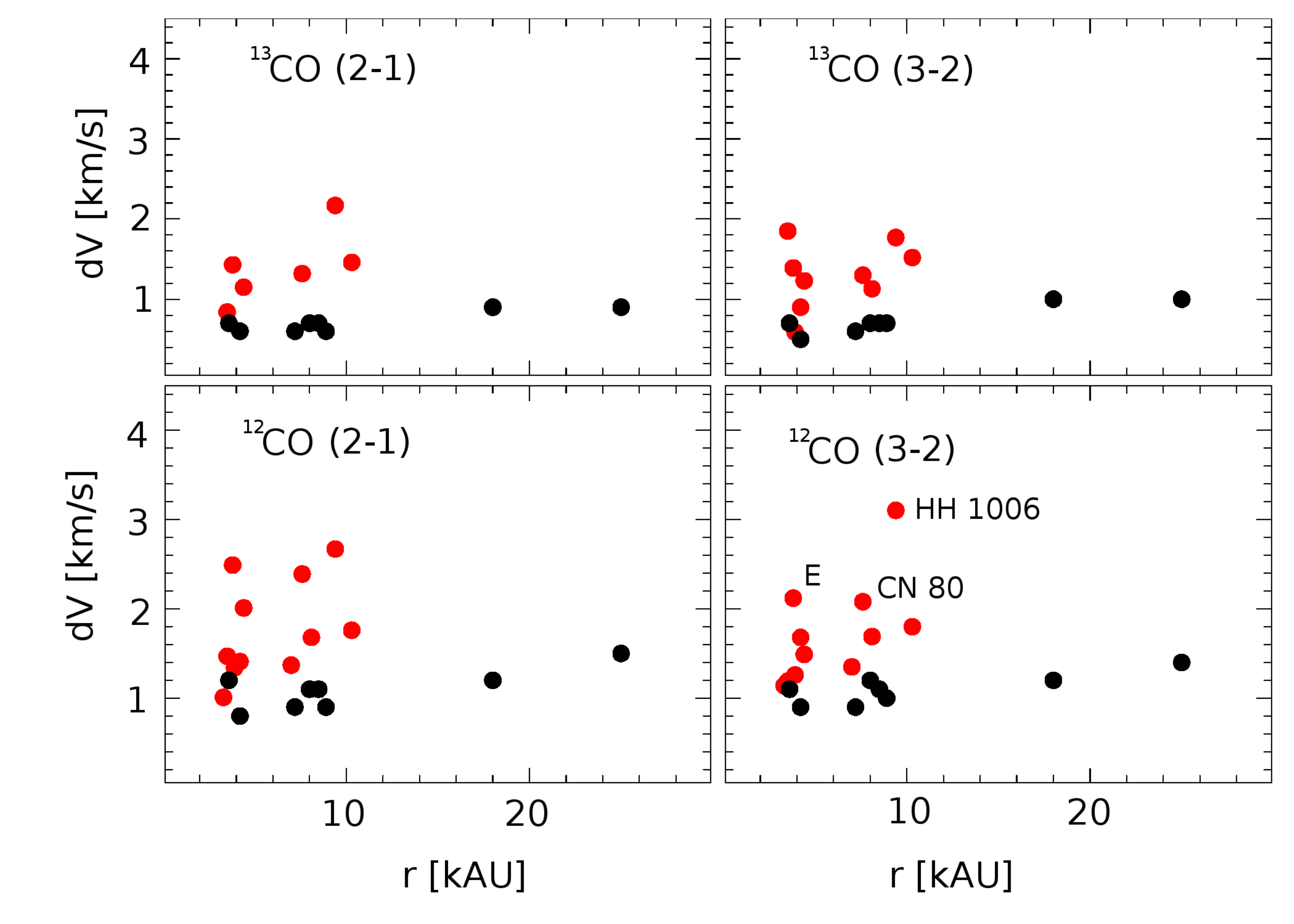}
\caption{Comparison of the half widths of the (2--1) and (3--2) lines in \twco and \thco of globulettes in the Carina and Rosette nebulae as a function of the globulette radius. The globulettes in Carina are plotted in red and those in the Rosette nebula are plotted in black. The three most deviating Carina globulettes are identified in the last panel.}
\label{fig:dv}
\end{figure}

Another difference is that the line widths of the CN objects are larger in all transitions than those of the RN objects. The observed line width depends neither on the distance to a source nor on the beam filling factor. Thus it is possible to compare directly the observed line widths observed for the CN and RN globulettes. The corresponding line half widths are plotted as a function of the globulette radius in kAU in Fig. \ref{fig:dv}, and  the mean, standard deviation, and median of the line widths observed for the two regions are listed in Table \ref{table:widths}. Objects  CN\,93-Frag (irregular), E (blended), and HH1006 (outflow) are not included. The line widths for the CN objects are significantly broader than for the RN objects. The differing line intensity ratios and line widths implicate that the density and temperature structure of the globulettes in the two regions are different. The CN globulettes are not only physically smaller and denser than those in RN, but have a different internal structure.

\begin{table}
  \caption[]{Line widths for Carina and Rosette globulettes.}
   \label{table:widths}
\begin{tabular}{lccccc}
\hline
\hline
& \twcop\,(3--2) &  \twcop\,(2-1)& \thcop\,(3--2)  & \thcop\,(2--1)   \\
& [\kmps] &  [\kmps]& [\kmps]& [\kmps]\\
\hline
          &     &  Carina   &     &     \\
$\Delta v$    & 1.49& 1.47& 1.18& 1.15\\
$\sigma$      & 0.27& 0.32& 0.49& 0.31\\
median        & 1.49& 1.41& 1.13& 1.15\\
              &     & Rosette    &     &     \\
$\Delta v$    & 1.10& 1.10& 0.70& 0.70\\
$\sigma$      & 0.17& 0.22& 0.18& 0.12\\
median        & 1.10& 1.09& 0.74& 0.70\\
\hline
\end{tabular}                                          
\end{table}

Excluding objects with bright rims there is a rather linear relation between the integrated intensity of the $^{12}$CO(2--1) and (3-2) lines and the optical mass as listed in Table~\ref{table:spectra}. However, the mass of objects with bright rims can be estimated from this intensity-mass relation. Clearly, they contain much more mass than the lower limit set from optical data. For the objects $j$ and F we infer masses of $>$~200 $M_{J}$ and for E $\sim$~100 $M_{J}$. The molecular lines in HH 1006 are broader than in other objects, which can be related to an outflow as discussed in Sect.~\ref{sec:velocity}. This object falls below the slope in the intensity-mass relation, which we conclude is related to large optical depth in the (2-1) transitions. Its mass amounts to 350 $M_{J}$ according to Sahai et al. (\cite{sahai12}) as derived from several higher transitions in CO, HCO$^{+}$, and HCN.

\section{Conclusions}
\label{sec:conclusions}

We present results based on APEX observations of (3--2) and (2--1)  $^{12}$CO and $^{13}$CO lines of molecular clumps in the Carina nebula. Most of these are globulettes listed in the optical survey by Grenman \& Gahm (\cite{grenman14}; G14), who concluded that on the whole these globulettes are smaller and denser than those found in other \ion{H}{ii} regions surrounding young stellar clusters. Aside from the globulettes, we also observed selected positions in the molecular shell associated with the \ion{H}{ii} region. A sample of 12 of the most massive objects, mainly from the G14 survey, were selected for observations. Molecular line emission was detected from all objects, and the results were compared with those from a survey by Gahm et al. (\cite{gahm13}) of radio emission from globulettes in the Rosette nebula. Below follows an account of the main results:

* Strong emission from background molecular clouds are present in the velocity interval -20 to +10 km s$^{-1}$ (LSR) at most positions, but the globulettes and associated foreground shells move at significantly more negative radial velocities and with a small internal spread in velocity including some very isolated objects located far from any shell structures. The globulettes are slightly shifted in velocity relative to adjacent elephant trunks and shell structures.

* A number of globulettes are located along an extended dust shell in the foreground of the nebulosity surrounding the Trumpler 14 cluster. These globulettes, and the associated shell, are blue-shifted by a few km~s$^{-1}$ relative to the mean velocity of the cluster in line with predictions from current numerical model simulations. However, in the same system the orientation of elephant trunks rooted in the shell and elongated globulettes indicate that the objects in question are influenced and sculpted by the interaction from massive stars in the central part of the Carina nebula, including the cluster Trumpler 16. 

* The line widths measured for the Carina globulettes are 50\% larger than those measured for Rosette objects and even larger for the non-typical objects HH1006, CN93frag, and F.  The HH 1006 exceptionally broad lines relate to the presence of a molecular outflow driven by an imbedded infrared star. The  $^{12}$CO(3--2)/$^{13}$CO(3--2) and $^{12}$CO(2--1))/$^{13}$CO(2--1) line ratios are also different for the Carina objects compared to the Rosette objects. For the Carina objects the integrated intensities of the (3--2) lines dominate over the (2--1) lines unlike in the Rosette nebula, where they are similar. The Carina nebula is at a larger distance, however, and for a similar mass the Carina globulettes appear smaller resulting in a smaller filling factor. Because of this circumstance we were unable to make a more detailed modelling of the structure of the Carina objects as carried out for the Rosette objects. We conclude that the Carina globulettes have different density and temperature structures compared to those in the Rosette nebula based on differences in line widths and relative line intensities, and we confirm the high average densities found in the optical survey.

* The close association between globulettes and shells/trunks, both in terms of location and velocity, indicates that the globulettes are detached remnants from eroding shell structures.

\begin{acknowledgements}
This work was supported by the Magnus Bergvall Foundation. M.M. acknowledges the support from the Finnish Graduate School in Astronomy and Space Physics. C.M.P. acknowledges generous support from the Swedish National Space Board. This research has made use of the products of the NASA/ESA Hubble Space Telescope, obtained at the Space Telescope Science Institute.

\end{acknowledgements}

\end{document}